\documentstyle[multicol,aps,epsfig]{revtex}

\begin{document}
%%%%%%%%%%%%%%%%%%%%%%%%%%%%%%%%%%%%%%%%%%%%%%
%%%%%%%%%%%%%%%%%%%%%%%%%%%%%%%%%%%%%%%%%%%%%%
\def\dd{{\rm d}}
\def\bea{\begin{eqnarray}}
\def\eea{\end{eqnarray}}
\def\bqo{\begin{quote}}
\def\eqo{\end{quote}}
\def\bc{\begin{center}}
\def\ec{\end{center}}
\def\om{\omega}
\def\xs{\xi^s}
\def\ep{\epsilon}
\def\vaep{\varepsilon}
\def\re{{\rm Re}}
\def\im{{\rm Im}}
\def\cii{\chi_{\rm{imp}}}
\def\ciif{\chi_{\rm{imp1}}}
\def\ciis{\chi_{\rm{imp2}}}
\def\tp{\tilde{p}}
\def\tw{\tilde{\omega}}
\def\Res{{\rm Residues}}
\def\sgn{{\rm sgn}}
\def\ip{ip_{n}}
\def\nonu{\nonumber}
%%%%%%%%%%%%%%%%%%%%%%%%%%%%%%%%%%%%%%%%%%%%%%%%%%
%\twocolumn[\hsize\textwidth\columnwidth\hsize\csname@twocolumnfalse\endcsname
%%%%%%%%%%%%%%%%%%%%%%%%%%%%%%%%%%%%%%%%%%%%%%%%%%

\title{
Korringa ratio of ferromagnetically correlated impure metals }

\author{Han-Oh Lee and Han-Yong Choi}

\address{Department of Physics, Institute for Basic Science Research,
and BK21 Physics Research Devision,\\
Sung Kyun Kwan University, Suwon 440-746, Korea.}

\date{\today}

\maketitle

\begin{abstract}
The Korringa ratio, $\cal K$, obtained by taking an appropriate combination of
the Knight shift and nuclear spin-lattice relaxation time, is calculated
at finite temperature, $T$, in the three-dimensional electron gas model,
including the electron-electron interaction, $U$, and non-magnetic impurity scatterings.
$\cal K$ varies in a simple way with respect to $U$ and $T$; it
decreases as $U$ is increased
but  increases as $T$ is raised.
However, $\cal K$ varies in a slightly more complicated way with respect to
the impurity scatterings;
as the scattering rate is increased,
$\cal K$ increases for small $U$ and low $T$, but  decreases for large $U$ or high $T$ regime.
This calls for a more careful analysis when one attempts to estimate the Stoner factor
from $\cal K$.

\end{abstract}


%\begin{multicols}{2}

\section{Introduction}
%%%%%%%%%%%%%%%%%%%%%%%%%%%%%%%%%%%%%%%%%%

The Korringa ratio, ${\cal{K}}$, given by $\cal{K}$$ = 1 /(T_1 T K^2)$, where $T_1$ is the nuclear
spin-lattice relaxation time, $K$ the Knight shift, and $T$ is the temperature,
is a useful concept \cite{korringa},
in that when $\cal{K}$ of a material is larger (smaller) than unity,
it is interpreted as a strong indication that the material
is antiferromagnetically (ferromagnetically) correlated.
Moriya, some forty years ago, showed that $\cal{K}$ is decreased when there is
electron-electron interaction, $U$, in a three-dimensional (3D) free electron gas \cite{moriya}.
Underlying physics is that $K^2$ increases more rapidly than $1/T_1 T$ for a ferromagnetically
correlated system as $U$ is increased.
It was later extended to include the non-magnetic impurity scattering effects by Shastry and Abrahams
\cite{shastry}. They showed that disorder enhances the Korringa ratio so that the impurity effects should
be included in addition to the electron-electron interaction when one analyzes the ratio, eg, when one
wants to estimate the Stoner factor of a material from $\cal{K}$.
These calculations were done at zero temperature.
Since experiments are performed at finite temperature,
we wish to extend their calcuations to the finite temperature, and understand
how $\cal{K}$ is changed as the temperature, $T$, the electron-electron interaction, $U$, and the impurity
scattering rate, $1/\tau$, are varied.

$K$ and $1/T_1 $, which determine $\cal{K}$, can be expressed in terms of
the spin susceptibility, $\chi({\bf q},\omega)$, of the conduction electrons.
Fulde and Luther calculated the impurity effects on the spin susceptibility
of almost ferromagnetic metals by including
isotropic non-magnetic impurity scatterings
and Dirac $\delta$-function electron-electron interaction at zero temperature,
in the small frequency and the small wavevector limit \cite{fulde}.
Their result was utilized by Shastry and Abrahams to calculate $\cal K$
of almost ferromagnetic dirty metals.
In the present work, we reformulate the spin susceptibility
with the Matsubara Green's function to consider the finite temperature effects
on the Korringa ratio at the presence of the electron-electron interaction and impurity
scatterings. The calculated $K$, $T_1$, and $\cal K$ agree in most cases with those of Shastry and Abrahams:
both $K$ and $1/T_1 T$ increase when $U$ or $1/\tau$ is increased, and $\cal K$ decreases
as $U$ is increased.
One interesting observation, different from their results, is that $\cal K$ increases as $1/\tau$ is increased
only when $both$ $U$ and $T$ are small. When $U$ is comparable with $U_{cr}$, or $T$ is with $\epsilon_F$,
$\cal K$ decreases as $1/\tau$ is raised.
$U_{cr}$ is the critical value of the electron-electron interaction at which the ferromagnetic instability
occurs, and $\epsilon_F$ is the Fermi energy.
Consequently, $\cal K$ vs. $U$ curves for several different $1/\tau$ values, at a fixed $T$,
cross each other. These crosses
also occur in $\cal K$ vs. $T$ at a given $U$ among different values of $1/\tau$.
It has been believed, since the Shastry and Abrahams work, that the impurity scatterings enhance $\cal K$
while the electron-electron interaction suppresses it.
The present result that impurity scatterings can enhance or suppress depending on $U$ or $T$ requires that
the above interpretation should be modified.
This calls for a more careful analysis of experimental data when one wishes to estimate
the strength of ferromagnetic correlation of a material.

The remainder of the present paper is organized as follows:
In Sec. II, we present the Matsubara formalism to include
the electron-electron interaction and impurity scatterings in calculating
$\chi({\bf q},\omega)$ at finite temperature.
This was carried out in the Matsubara frequency and then analytically
continued to the real frequency.
In Sec. III, $K$, $1/T_1 T$, and $\cal K$ are calculated by numerically integrating the
obtained equations, as $U$, $1/\tau$, and $T$ are varied. These results are compared
with the free electron gas model as a reference, and with previously reported calculations.
Finally, Sec. IV is for the summary and concluding remarks.

%%%%%%%%%%%%%%%%%%%%%%%%%%%%%%%%%%%%%%%%%%%%
\section{Formalism}\label{s2}
%%%%%%%%%%%%%%%%%%%%%%%%%%%%%%%%%%%%%%%%%%%%
%\hspace{0.3cm}

The Korringa ratio is obtained from the Knight shift and nuclear spin-lattice relaxation time
as $\cal{K} $$= 1/(T_1 T K^2)$, and $K$ and $1/T_1$ are given in terms of the spin
susceptibility as
\bea
K \propto \lim_{{\bf q} \rightarrow 0 \atop \omega \rightarrow 0} Re \chi({\bf q},\omega),
\nonumber \\
\frac{1}{T_1 T} \propto \lim_{\omega \rightarrow 0} \sum_{{\bf q}}
\frac{1}{\omega} Im \chi({\bf q},\omega).
\eea
Calculating the Korringa ratio, therefore, amounts to calculation of $\chi({\bf q}, \omega)$
including the electron-electron interaction and the impurity scatterings at finite temperature.
This can most conveniently be done in terms of the Matsubara Green's function in the
imaginary frequency to obtain $ \chi({\bf q}, i\omega)$. It is then analytically continued
to yield $\chi({\bf q},\omega) $ in the real frequency by substituting $i\omega \rightarrow
\omega +i\delta$, where $\delta$ is a positive infinitesimal.
We, therefore, reformulate
the Fulde and Luther result
with the Matsubara method for spin susceptibility and
construct $K$ and $(T_1 T)^{-1}$.
The spin susceptibility including the vertex corrections
from the Coulomb interaction and electron-impurity scatterings
can be expressed in terms of the Feynman diagram as in Fig.\ 1(a).
Here, the solid lines denote
the renormalized electron Green's function $G({\bf k},ip)$, and
$\Gamma$ denotes the vertex function, where
$ip$ and $i\omega$ are, respectively,  the fermion and boson Matsubara frequencies,
and $\bf k$ and $\bf q$ are the correspinding wavevectors.
$\Gamma$ includes both the impurity and the Coulomb interaction as shown in Fig. 1(b), but $G$
does not include the renormalization from the electron-electron interaction
as in the previous works \cite{fulde,shastry}.
The neglect of $U$ in calculating $G$ may be justified because Schrieffer and Berk
showed that this model reproduces the long wave length static susceptibility which
would be obtained from a more complete calculation including the renormalizations
due to $U$ \cite{schrieffer}.

$\Gamma$, represented in Fig.\ 1(b),
can be written as
%%gamma%%%%%%%%%%%%%%%%%%%%%%%%%%%%%%%%%%%%%
\bea\label{gamma}
\Gamma({\bf q}, ip, i\omega) =
1&-& \frac{U}{\beta}\sum_{ip'} \sum_{{\bf k}'}\, G({\bf k}', ip')\,
G({\bf k'+q}, ip'+i\omega)\, \Gamma({\bf q}, ip', i\omega)\nonumber\\
&+& \frac{1}{2 \pi N \tau}
\sum_{{\bf k}'}\, G({\bf k}', ip)\, G({\bf k'+q}, ip + i\omega)\,
\Gamma({\bf q}, ip, i\omega) .
\eea
%%%%%%%%%%%%%%%%%%%%%%%%%%%%%%%%%%%%%%%%%%%
Here, $\beta =1/k_B T$,
$N$ the single-spin density of states (DOS) at the Fermi level,
and $G({\bf k},ip)=1/[ip +\frac{i}{2\tau} \sgn(p)-\xi_{\bf k}]$,
where $\sgn(p)$ is $1$ if $p$ is positive or zero and is $-1$ otherwise,
and $\xi_{\bf k}$ is energy measured from the Fermi surface,
or $\xi_{\bf k} = \hbar^2{\bf k}^2/2m-\ep_F$.
$\chi$ can be written as in Ref.\cite{fulde}
in Matsubara form as
%%%%%%%%%%%%%%%%%%%%%%%%%%%%%%%%%%%%%%%%%%%%
%%Chi%%%%%%%%%%%%%%%%%%%%%%%%%%%%%%%%%%%%%%%
\bea\label{r1}
\chi ({\bf q},i\omega)&=&-\mu_{B}^{2}
\frac{2}{\beta} \sum_{ip} \sum_{{\bf k}}\,
G({\bf k},ip)\, G({\bf k+q},ip +i\omega)\,\Gamma ({\bf q},ip, i\omega)\;\;
\\\nonu
&=&2\mu_{B}^{2}N
\frac{\cii({\bf q},i\om)}{1-NU\cii({\bf q},i\om)} ,
\eea
%%Ximp%%%%%%%%%%%%%%%%%%%%%%%%%%%%%%%%%%%%%
where
\bea
\cii({\bf q}, i\omega) = -\,\frac{2\pi}{\beta}\sum_{ip}
\frac{J({\bf q}, ip, i\omega)}{1 - \frac{1}{\tau}\,
J ({\bf q}, ip, i\omega)}
\eea
and
\bea
J({\bf q}, ip, i\omega)&=&\frac{1}{2\pi N} \sum_{{\bf k}}\,
G({\bf k}, ip)\,G({\bf k+q}, ip+i\omega)\\\nonu
&=&
\frac{im^{2}}{4\pi^{2} N q}
\ln \left[ \frac
{\sgn(\tilde{p}) \sqrt{i\tilde{p}\!+\!\mu} +
 \sgn(\tilde{p}\!+\!\tilde{\omega})
     \sqrt{i(\tilde{p}+\tilde{\omega})\!+\!\mu}
      +\sqrt{\ep_q}}
{\sgn(\tilde{p})\sqrt{i\tilde{p}\!+\!\mu} +
 \sgn(\tilde{p}\!+\!\tilde{\omega})
     \sqrt{i(\tilde{p}\!+\!\tilde{\omega})\!+\!\mu}
     -\sqrt{\ep_q}} \right] .
\eea
%%%%%%%%%%%%%%%%%%%%%%%%%%%%%%%%%%%%%%%%%%%%%
Here, we put $k_B=\hbar=1$,
$\mu_B$ the Bohr magneton, $m$ the electron mass,
$\ep_q=q^2/2m$, $i\tilde{p}=ip+\frac{i}{2\tau}\sgn(p)$,
and $\mu$ is the chemical potential.
In order to obtain $\chi$ in the real frequency,
analytic continuation of $i\om\to \om+i\delta$
is performed after the frequency summation.
The frequency summation can be done in complex plane
using contour integral,
where summation is replaced by inegrals
over branch cut lines\cite{mahan}.
Then, $\cii$ can be written as
%%Ximp-ac%%%%%%%%%%%%%%%%%%%%%%%%%%%%%%%%%%%%%
\bea\label{acf}
\cii(q,\om)&&=
-i\int_{-\infty}^{\infty}\! d\varepsilon\!
\biggm\{
\bigl[ n_{F}(\varepsilon + \omega) - n_{F}(\varepsilon)\bigr]
\,F(\varepsilon -\frac{i}{2\tau},
  \varepsilon + \omega +\frac{i}{2\tau}) \nonumber\\
&&+n_{F}(\varepsilon)
\,F(\varepsilon + \frac{i}{2\tau},
  \varepsilon + \omega + \frac{i}{2\tau})
  -n_{F}(\varepsilon+\omega)
\,F(\varepsilon -\frac{i}{2\tau},
  \varepsilon + \omega - \frac{i}{2\tau})
\biggm\},
\eea
%%%%%%%%%%%%%%%%%%%%%%%%%%%%%%%%%%%%%%%%%%%
where we introduce for covenience an expression
%%F-ac%%%%%%%%%%%%%%%%%%%%%%%%%%%%%%%%%%%%%
\bea
F(\varepsilon \pm \frac{i}{2\tau},\varepsilon + \omega \pm
\frac{i}{2\tau})\,=\, \frac{J(\varepsilon \pm
\frac{i}{2\tau},\varepsilon + \omega \pm
\frac{i}{2\tau})}{1-\frac{1}{\tau}\,J(\varepsilon \pm
\frac{i}{2\tau},\varepsilon + \omega \pm
\frac{i}{2\tau})} ,
\eea
%%%%%%%%%%%%%%%%%%%%%%%%%%%%%%%%%%%%%%%%%%%
and
%%J-ac%%%%%%%%%%%%%%%%%%%%%%%%%%%%%%%%%%%%%
\bea\label{ko5}
J(\vaep\pm\frac{i}{2\tau},\vaep+\omega\pm
\frac{i}{2\tau})
=\frac{im^{2}}{4\pi^{2}Nq}
\ln \!\left(
\frac{\pm \sqrt{\vaep+\mu \pm \frac{i}{2\tau}} \pm
\sqrt{\vaep+\omega+\mu \pm \frac{i}{2\tau}} + \sqrt{\ep_q}}
{\pm \sqrt{\vaep+\mu \pm \frac{i}{2\tau}} \pm
\sqrt{\vaep+\omega+\mu \pm \frac{i}{2\tau}} - \sqrt{\ep_q}}
\right) .\nonumber\\ \;
\eea
%%%%%%%%%%%%%%%%%%%%%%%%%%%%%%%%%%%%%%%%%%%%
Here, $n_F (\epsilon) = 1/[1 +\exp(\beta \epsilon)]$ is the Fermi distribution function.

Now we can express $K$ and $(T_1 T)^{-1}$ with this susceptibility,
through taking its real and imaginary parts.
$(T_1 T)^{-1}$ can be written as
%%%%%%%%%%%%%%%%%%%%%%%%%%%%%%%%%%%%%%%%%%%%
\bea\label{t1t}
\frac{1}{T_{1}T}
\propto \frac{\mu_{B}^{2}N}{\pi^{2}}\int_{0}^{\infty}\!\!dq\, q^{2}
\frac{\partial \ciis /\partial \omega (q,\omega \to
0)}{\left[1-NU\ciif(q,\omega \to 0) \right]^{2}}
\eea
%%%%%%%%%%%%%%%%%%%%%%%%%%%%%%%%%%%%%%%%
where $\chi_2$ and $\ciis$ denote
the imaginary parts of $\chi$ and $\cii$, respectively.
$\chi_{\rm imp1}$ and $\chi_{\rm imp2}$ in Eq.\ (\ref{t1t}) can be written as
%%%%%%%%%%%%%%%%%%%%%%%%%%%%%%%%%%%%%%%%%%
\bea\label{tx1t}
\chi_{\rm imp1}(t)&=&
\int_{-\infty}^{\infty}dx\,
\left(\frac{1}{1+e^{Bx}}\right)
Im\! \left\{
\frac{J(x\!+\!i\eta,x\!+\!i\eta)}
{1-\eta J(x\!+\!i\eta,x\!+\!i\eta)}
\right\} ,\\
%%%%%%%%%%%%%%%%%%%%%%%%%%%%%%%%%%%%%%%%%
\label{tx2pt}
\frac{\partial \chi_{\rm imp2}(t)}{\partial \omega} \!&=&\!
\frac{1}{2 \epsilon_F} \int_{-\infty}^{\infty}\!\! dx
\frac{-Be^{Bx}}{(1+e^{Bx})^2}
Re\!\biggm[
\left\{\frac{J(x\!+\!i\eta,x\!+\!i\eta)}
{1-\eta J(x\!+\!i\eta,x\!+\!i\eta)}\right\}
-\left\{\frac{J(x\!-\!i\eta,x\!+\!i\eta)}
{1-\eta J(x\!-\!i\eta,x\!+\!i\eta)}\right\}
\biggm] ,
\eea
%%%%%%%%%%%%%%%%%%%%%%%%%%%%%%%%%%%%%%%%%%
where $B = \beta \ep_F$, $\eta=(2\tau\ep_F)^{-1}$,
$M=\mu/\ep_F$, $t=q/k_F$, $x=\vaep/\ep_F$, and
%%%%%%%%%%%%%%%%%%%%%%%%%%%%%%%%%%%%%%%%%%%
\bea
J(x\! \pm\! i\eta,x\!+\! i\eta)=\frac{i}{2t}
\ln\!\left(
\frac{\pm\sqrt{x\!+\!M\!\pm\! i\eta}+\!\sqrt{x\!+\!M\!+\!i\eta}+t}
{\pm\sqrt{x\!+\!M\!\pm \!i\eta}+\!\sqrt{x\!+\!M\!+\!i\eta}-t}
\right).
\eea
%%%%%%%%%%%%%%%%%%%%%%%%%%%%%%%%%%%%%%%%%%%
In the same way, $K$ is written as
\bea\label{ko6}
K& \propto &\lim_{\om\to0 \atop {\bf q} \to 0}\chi_1({\bf q},\om)\nonumber\\
% &=&2\mu_B^2 N\, \re\!\left[
%     \frac{\cii(0,0)}{1-NU\cii(0,0)}\right]\nonumber\\
 &=&2\mu_B^2 N\, Re\!\left[\frac{\ciif(0,0)}
 {1-NU\ciif(0,0)}\right],
\eea
%%%%%%%%%%%%%%%%%%%%%%%%%%%%%%%%%%%%%%%%%%%
where $\ciif(0,0)$ is given from Eq.\ (\ref{tx1t}) by
%%%%%%%%%%%%%%%%%%%%%%%%%%%%%%%%%%%%%%%%%%%
\bea\label{k}
\ciif(0,0) =  \int_{-\infty}^{\infty}\!dx\;
\frac{1}{1+e^{Bx}} Re\!\left[
\frac{1}{2\sqrt{x+M+i\eta}-i\eta} \right].
\eea
%%%%%%%%%%%%%%%%%%%%%%%%%%%%%%%%%%%%%%%%%%%%
$\eta J$ in the
denominater in Eqs.\ (\ref{tx1t}) and (\ref{tx2pt}),
and $i\eta$ outside the sqare root in Eq.\ (\ref{k})
are from the impurity vertex correction.
These vertex parts from Eqs.\ (\ref{tx1t}) and (\ref{k})
cause the crosses between the curves of $\cal{K}$ with $T$ and $U$.

%%%%%%%%%%%%%%%%%%%%%%%%%%%%%%%%%%%%%%%%%%%
\section{ results}\label{s3}
%%%%%%%%%%%%%%%%%%%%%%%%%%%%%%%%%%%%%%%%%%%

Let us first consider the free electron gas model for a reference
before we present the calculations for ferromagnetically correlated dirty metals.
For a 3D free electron gas model, it is simple to calculate the static susceptibility
$\chi({\bf q}, \omega\rightarrow 0)$ as a function of $T$ \cite{huang}.
From $\chi$, $K$ and $(T_1 T)^{-1}$ are given as
\bea
K &\propto& \sqrt{\frac{\mu}{\epsilon_F} } \left[ 1 -
\frac{\pi^2}{24} \left( \frac{T}{\mu} \right)^2 -\cdots \right],
\nonumber \\
\frac{1}{T_1 T} &\propto& 2\pi \mu_B^2 N^2 \frac{T}{\epsilon_F}
\ln\left( 1 +e^{\mu/T} \right),
\eea
where the chemical potential $\mu$ is given by
\bea
\mu \approx \left\{
\begin{array}{ll}
\epsilon_F \left[ 1 - \frac{\pi^2}{12} \left(
\frac{T}{\epsilon_F} \right)^2 \right], & {\rm for} ~T\ll \epsilon_F,
\\
T \ln \left[ \frac{4}{ 3\sqrt{\pi}}
\left(\frac{\epsilon_F}{T} \right)^{3/2} \right], & {\rm for}~ T \gg \epsilon_F .
\end{array}
\right.
\eea
We, therefore, have
\bea
\frac{K}{K_0} \approx \left\{
\begin{array}{ll}
1-\frac{\pi^2}{12} \left( \frac{T}{\epsilon_F} \right)^2,
& {\rm for}~T\ll \epsilon_F, \\
1/T, & {\rm for}~ T \gg \epsilon_F ,
\end{array}
\right.
\label{knight0}
\eea
and
\bea
\frac{(T_1 T)^{-1}}{(T_1 T)_0^{-1}} \approx \left\{
\begin{array}{ll}
1-\frac{\pi^2}{12} \left( \frac{T}{\epsilon_F} \right)^2,
& {\rm for}~T\ll \epsilon_F, \\
1/\sqrt{T}, & {\rm for}~ T \gg \epsilon_F ,
\end{array}
\right.
\label{nslr0}
\eea
where $K_0 = 2 \mu_B^2 N$ and $(T_1 T)_0^{-1} = 2\pi \mu_B^2 N^2$ denote, respectively,
their zero temperature values at $U=1/\tau=0$.
From these elementary considerations, we see that
\bea
\cal{K} \propto \left\{
\begin{array}{ll}
1 +\frac{\pi^2}{12} \left( \frac{T}{\epsilon_F} \right)^2 ,
& {\rm for}~T\ll \epsilon_F, \\
T^{3/2}, & {\rm for}~ T \gg \epsilon_F .
\label{kappa0}
\end{array}
\right.
\eea

Introducing impurity scatterings and electron-electron interaction does not
alter in an essential way
the $K$, $1/T_1 T$ and $\cal K$ vs.\ $T$ behavior of Eqs.\ (\ref{knight0}), (\ref{nslr0}), and (\ref{kappa0}),
but change their $T$ dependence by altering the three coefficients:
(a) the value at $T=0$, (b) the low $T$ coefficient proportional to $T^2$ (Lorenzian width),
and (c) the high $T$ coefficient,
as shown in Figs.\ 2 and 3.
$K/K_0$ as a function of $T$ for a set of $\eta = 1/(2 \tau \epsilon_F)$ and $ NU$, where
$NU_{cr} = 1$ is the critical value at which the ferromagnetic instability occurs
at $T=$ and $\eta=0$, are shown in Fig.\ 2(a).
Both $NU$ and $\eta$ enhance $K$ in agreement with Shastry and Abrahams.
The low temperature coefficients of $K/K_0$ vs.\ $T$  can be calculated analytically
as functions of $NU$ and $\eta$:
(a) The $T=0$ peak is given by $a/(1-aNU)$, and
(b) the low $T$ Lorentzian width by $2 \sqrt{1-aNU}/\sqrt{b}$,
where  $a=1+\frac{\pi}{4}\eta-\frac{3}{8}\eta^2$ and
$b=\frac{\pi^2}{12}[1-(1+\frac{\pi}{4})\eta+
\frac{1}{4}(3+\pi+\frac{\pi^2}{4})\eta^2]$.
The numerical calulations agree perfectly well
with the analytically obtained asymptotic behavior.
In Fig.\ 2(b), $(T_1 T)^{-1}/(T_1 T)_0^{-1}$ as a function of $T$ are shown
for a set of $\eta$ and $ NU$. Their general behavior is almost the same
with that of $K/K_0$.
In Ref.\ \cite{pennington} and \cite{tycko}, $(T_1 T)^{-1}$ measured as a function
of $T$ is almost constant above $T_c$ in the fullerene superconductors.
It is difficult, however, to compare the present results with experimental observations
because  the lattice constant changes from
the thermal expansion. The lattice constant expansion affects the DOS and,
consequently,  $K$ and $1/T_1 T$.

In Fig.\ 3(a), we show $\cal K$ as a function of $T$ for a set of $\eta$ and $ NU$. One can see that
(a) the $T=0$ value is suppressed by $NU$, but, by $\eta$, it is increased for small $U$
and decreased for large $U$, ($\cal K $ = 1  for $T=U=\eta=0$.)
(b) the low $T$ coefficient is nearly independent of $NU$ and $\eta$, and
(c) the high $T$ coefficient proportional to $T^{3/2}$ is reduced as $\eta$ becomes larger.
Above changes cause the $\cal K$ vs. $T$ curves cross each other as can be seen in Fig.\ 4.
These crosses do not appear if we neglect the impurity vertex correction in $\ciif$, that is,
if we neglect the $1/\tau$ term in the denominator of $\ciif$,
and their origin is this impurity vertex correction, as will be discussed in more detail below.
Now, let us consider the $T=0$ limit case, which can directly be compared with Shastry and Abrahams.
In Fig.\ 3(b), we show $\cal K$ as a function of $NU$ for $\eta = 10^{-4}, ~0.05$, and 0.1 at $T=0$.
The three $\cal K$ vs. $NU$ curves cross each other as $NU$ is increased, as in Fig.\ (a) discussed
above.
These crosses did not occur in Shastry and Abrahams, who neglected the impurity vertex correction in $\ciif$.
At $T=0$, and for small $q$ and $\om$, the impurity enhanced susceptibility, $\cii$, can be expanded
by integrating Eq.\ (\ref{acf}) as,
\begin{equation}
\chi_{\rm imp} (q,\omega) \approx \cii^{(0)}(q,\om)\frac{D_0q^2}{D_ 0q^2-i\om},
%+\frac{\pi}{4} \eta+ \frac{1}{8} \eta^2 ,
\end{equation}
where the diffusion constant is given by $D_0=\tau v_F^2/3$, and
$\cii^{(0)}(0,0)=\cii(0,0)=\ciif(0,0)=1+\frac{\pi}{4} \eta+ \frac{1}{8} \eta^2$,
up to the order of $\eta^2$. The
$\pi\eta/4$ in $\cii^{(0)}$ comes from the impurity vertex correction  in the $\om=q=0$ limit,
and $\eta^2/8$ from the self-energy correction due to impurity scatterings
in the renormalized Green's funtions.
Shastry and Abrahams neglected the impurity vertex corrections in calculating
$\chi_{\rm imp}(q,\omega=0)$, and found that $\chi_{\rm imp}^{(0)} (0,0) \approx 1 + \eta^2 /8$
in agreement with the present result.
These two terms enhance $\ciif(0,0)$ compared with the free electron value of 1.
Because $\ciif(q,0)$ is multiplied by $NU$ in the denomenator when calculating
$K$ or $1/(T_1 T)$ as in Eqs.\ (\ref{t1t}) or (\ref{ko6}), the effects of $U$ are stronger for larger $\eta$.
Consequently, $\cal{K}$ decreases faster  for larger $\eta$ as $NU$ is increased.
This, together with the fact that $\cal K$ is larger for larger $\eta$ at $U=0$, causes the crosses
among the curves of $\cal K$ vs. $NU$ as shown in Fig.\ 3(b)
Since the $\pi\eta/4$ term in $\ciif(0,0)$ is dominant contribution from the impurity scatterings
compared with $\eta^2/8$,
these crosses do not occur without the impurity vertex correction.
It should be interesting if this
prediction can be confirmed by experiments.

%%%%%%%%%%%%%%%%%%%%%%%%%%%%%%%%%%%%%%%%%%%%
\section{Summary and concluding remarks}
%%%%%%%%%%%%%%%%%%%%%%%%%%%%%%%%%%%%%%%%%%%%
In this paper,
we studied the Korringa ratio ${\cal K }= 1/(T_1 T K^2)$, Knight shift $K$, and nuclear spin-lattice
relaxation time $T_1$ of ferromagnetically correlated
impure metals at finite temperature.
They were calculated from the spin susceptibility including the self-energy correction
due to the impurity scatterings,
and the vertex correction due to the $\delta$-function electron-electron
interaction and impurity scatterings.
The spin susceptibility was formulated with Matsubara Green's function
to consider the finite temperature effects and analytically continued to the real frequency.
The $K$ and $(T_1 T)^{-1}$ decrease with $T$, but
$\cal{K}$ increases as $T$ is raised because $K^2$ decreases more rapidly than
$(T_1 T)^{-1}$ with $T$.
Also, $K$ and $(T_1 T)^{-1}$ increase as $U$ or $\eta$ is increased,
and $\cal{K}$ decreases as $U$ is increased.

The conventional interpretation of the Korringa ratio, as mentioned in Introduction, is that
when $\cal{K}$ of a material is larger (smaller) than unity,
it implies that the material
is antiferromagnetically (ferromagnetically) correlated.
The Shastry and Abrahams contribution is that $\cal K$ is increased by the non-magnetic
impurity scatterings, so that $\cal K$ of ferromagnetically correlated dirty metals,
where $\cal K$ is suppressed by the ferromagnetic correlation but increased by the impurity
scatterings, can be close to 1, just like the free electron case.
This means that one should be careful in extracting the strength of ferromagnetic correlation
from the Korringa ratio.

Our observation in the present work is that the Korringa ratio is increased by the non-magnetic
impurity scatterings when $both$ $U$ and $T$ are small, but $\cal K$ is decreased by the
impurity scatterings when $U$ is comparable with $U_{cr}$ or $T$ is comparable with $\epsilon_F$.
This implies that $\cal K$ is enhanced by the impurity scatterings only for
weakly ferromagnetic dirty metals at low $T$, but supressed when $U$ or $T$ is large.
This causes the crosses
among the $\cal{K}$ vs. $NU$ curves for the
different impurity scattering rates $1/\tau$ in the present calculations.
The crosses also occur in $\cal{K}$ vs. $T$ at a given $U$ among differnent values of $1/\tau$.
Those crosses are caused by
the impurity vertex correction in $\ciif(0,0)$,
which was neglected in the previous calculations.
The present exact numerical calculations within the finite temperature extension
of the Fulde and Luther formulation, shown in Fig.\ 3,
confirm that above behavior is due to the impurity vertex corrections.
This observation requires that the interpretation according to Shastry and Abrahams
should be modified, and one should do more careful analysis
in estimating the strength of ferromagnetic correlation of impure materials
from the Korringa ratio.

We would like to thank Hyun-Cheol Lee for useful conversations.
This work was supported by Korea Science \& Engineering
Foundatoin through Grant No.\ 1999-2-114-005-5, and through
Intern Program No.\ 99-0200-0002-3 (HOL), and by Ministry of
Education through Brain Korea 21 SNU-SKKU Physics Program.

%%%%%%%%%%%%%%%%%%%%%%%%%%%%%%%%%%%%%%%%%%%%%%

%%%%%%%%%%%%%%%%%%%%%%%%%%%%%%%%%%%%%%%%%%%%%%

\begin{figure}
\epsfig{figure=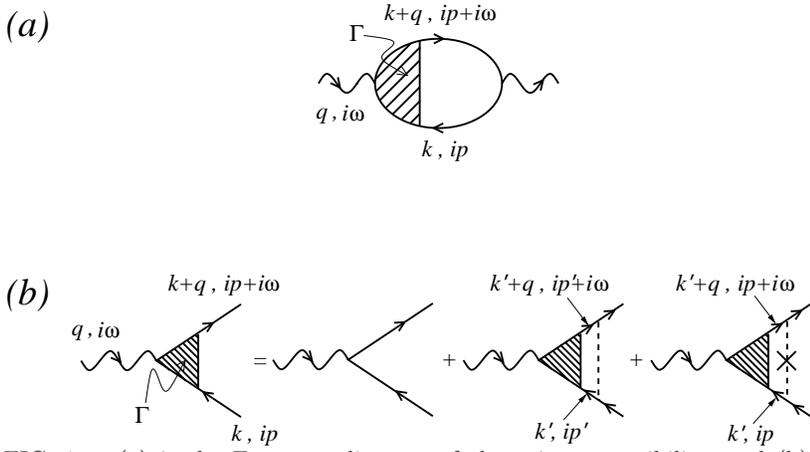,width=0.6\linewidth}
\caption{
(a) is the Feynman diagram of the spin susceptibility, and
(b) is for the renormalized vertex $\Gamma$ including the impurity and Coulomb interactions.
The solid line  indicates the electron Green's function, the dashed line the electron-electron interaction,
and the dashed line with t a cross stands for the impurity scatterings.
$ip$ and $i\omega$ are, respectively, the fermion and boson Matsubara frequencies,
and $k,q$ are correspinding wave vectors.
\label{fig1} }
\end{figure}

\vspace{1in}

\begin{figure}
\epsfig{figure=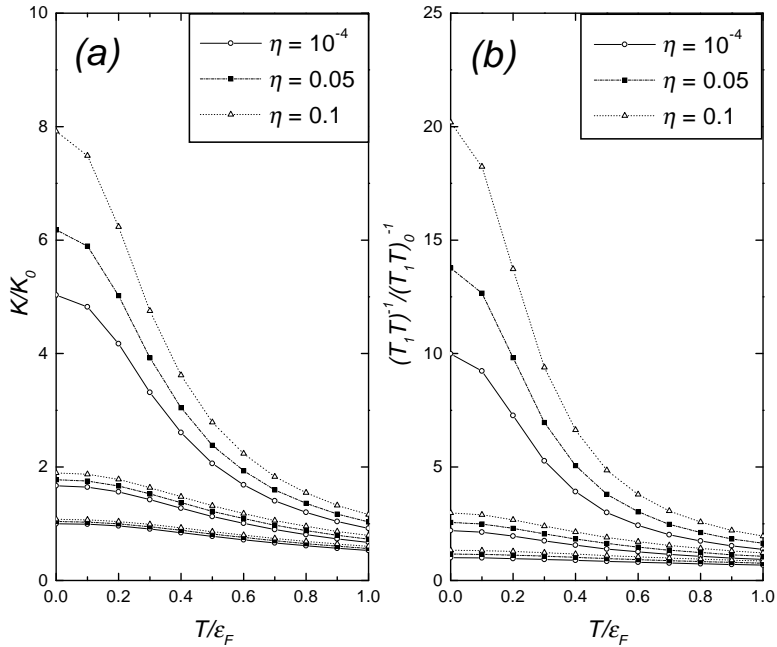,width=0.6\linewidth}
\caption{
(a) is the normalized Kight shift $K/K_0$ vs. the
reduced temperature, $T/\epsilon_F$, and (b) is $(T_1 T)^{-1}/{(T_1 T)_0}^{-1}$ vs. $T/\epsilon_F$.
$K_0$ and ${(T_1 T)_0}^{-1}$ are, respectively, the zero temperature values of $K$
and $(T_1 T)^{-1}$ at $T=U=\eta=0$ limit. $NU=$ 0, 0.4, 0.8 from bottom to top and
$\epsilon_F = 0.25$ eV.
$K/K_0$ and $(T_1 T)^{-1}/{(T_1 T)_0}^{-1}$ decrease more rapidly with $T$
when $NU$ or $\eta$ is large.
\label{fig2} }
\end{figure}

\begin{figure}
\epsfig{figure=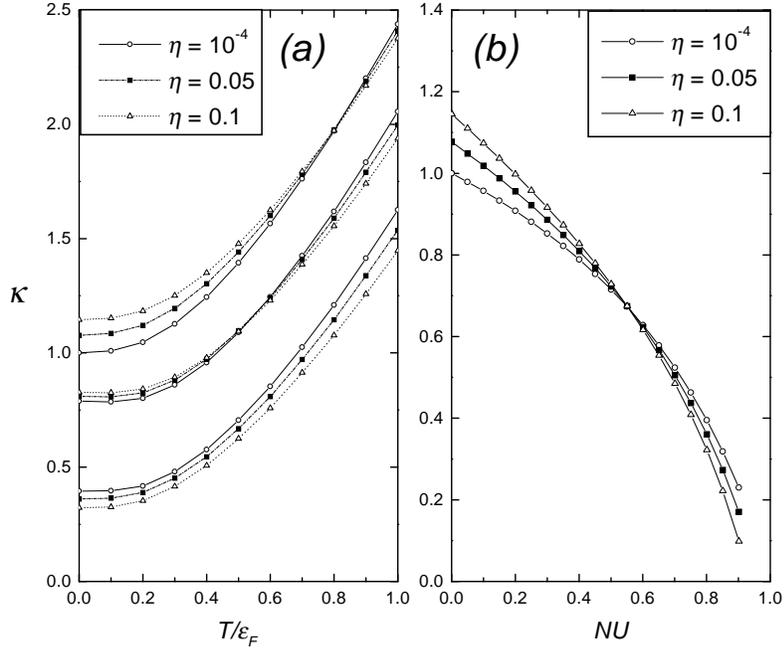,width=0.6\linewidth}
\caption{ (a) is the Korringa ratio $\cal{K}$
vs. $T/\epsilon_F$, and (b) is $\cal{K}$ vs. $NU$ at $T=0$ limit.
$\cal{K}$=1 in the $T=U=\eta=0$ limit. $\epsilon_F = 0.25$ eV, and $NU=$ 0, 0.4, 0.8 from bottom to top
for Fig.\ (a).
The crosses among the curves for different $\eta$ occur
at lower $T$ as $NU$ is increased,
implying that in $\cal{K}$ vs. $NU$ crosses occur
at lower $NU$ when $T$ is increased.
Fig.\ (b) can be compared with Ref.\ 3, where there are no crosses among
different $\eta$ values.
\label{fig3} }
\end{figure}

%%%%%%%%%%%%%%%%%%%%%%%%%%%%%%%%%%%%%%%%%%%%%%

%\end{multicols}

\end{document}